\documentclass{agujournal2019}
\usepackage{url} 
\usepackage{lineno}
\usepackage{amsmath}
\usepackage[inline]{trackchanges} 
\usepackage{soul}
\usepackage{amsmath, amsfonts, amssymb}
\usepackage{xr}

\usepackage{hyperref}
\hypersetup{breaklinks = true,
	        colorlinks = true,
	        linkcolor = blue,
            urlcolor = blue,
	        citecolor = blue}

\newcommand{\bu}{\boldsymbol{u}}
\newcommand{\btau}{\boldsymbol{\tau}}

\newcommand{\bnabla}{\boldsymbol{\nabla}}
\newcommand{\JpbH}{\bzhat \bcdot (\bnabla p_b \times \bnabla H)}

\newcommand{\badv}{\boldsymbol{\mathcal{A}}} 
\newcommand{\bcalU}{\boldsymbol{\mathcal{U}}} 
\newcommand{\bvisc}{\boldsymbol{\mathcal{V}}} 
\newcommand{\bcdot}{\boldsymbol{\cdot}}

\newcommand{\bxhat}{\boldsymbol{\hat{x}}}
\newcommand{\byhat}{\boldsymbol{\hat{y}}}
\newcommand{\bzhat}{\boldsymbol{\hat{z}}}
\renewcommand{\equiv}{\ensuremath{\stackrel{\mathrm{def}}{=}}}

\draftfalse

\journalname{Journal of Geophysical Research: Oceans}

\begin{document}
\justify

%
%


\title{Asymmetric response of the North Atlantic gyres to the North Atlantic Oscillation}

%
%

\authors{Dhruv~Bhagtani\affil{1,2}\thanks{142 Mills Road, Acton, 2601, ACT, Australia},
Andrew~McC.~Hogg\affil{1,2,3},
Ryan~M.~Holmes\affil{4},
Navid~C.~Constantinou\affil{3,5}, and
Hemant Khatri\affil{6}}

\affiliation{1}{Research School of Earth Sciences, Australian National University, Canberra, ACT, Australia}
\affiliation{2}{ARC Center of Excellence for Climate Extremes, Australia}
\affiliation{3}{ARC Center of Excellence for the Weather of the 21st Century, Australia}
\affiliation{4}{Australian Bureau of Meteorology, Sydney, NSW, Australia}
\affiliation{5}{School of Geography, Earth and Atmospheric Sciences, University of Melbourne, Parkville, VIC, Australia}
\affiliation{6}{Department of Earth, Ocean, and Ecological Sciences, University of Liverpool, United Kingdom}

\correspondingauthor{Dhruv Bhagtani}{dhruv.bhagtani@princeton.edu}

\begin{keypoints}
\item Both the subtropical and subpolar gyres strengthen when the North Atlantic Oscillation (NAO) phase is positive.
\item NAO-induced winds drive 90\% of the circulation anomalies in the subtropical gyre's interior.
\item Surface wind stress, non-linear advection, surface heating, and topography steer NAO-induced anomalies in the subpolar gyre.
\end{keypoints}

\begin{abstract}
The North Atlantic Oscillation (NAO) is a leading mode of atmospheric variability, affecting the North Atlantic Ocean on sub-seasonal to multi-decadal timescales.
The NAO changes the atmospheric forcing at the ocean's surface, including winds and surface buoyancy fluxes, both of which are known to impact large-scale gyre circulation.
However, the relative role of other physical processes (such as mesoscale eddies and topography) in influencing gyre circulation under NAO variability is not fully understood.
Here, we analyze a series of ocean--sea ice simulations using a barotropic vorticity budget to understand the long-term response of the North Atlantic gyre circulation to NAO forcing.
We find that for each standard deviation increase in the NAO index, the subtropical and subpolar gyres intensify by $0.90$~Sv and $3.41$~Sv (1$\,\mathrm{Sv}$ $\equiv 10^{6}\,\mathrm{m}^{3}\,\mathrm{s}^{-1}$) respectively.
The NAO-induced wind stress anomalies drive approximately 90\% of the change in the subtropical gyre's interior flow.
However, in the subpolar gyre's interior, in addition to wind stress, flow-topography interactions, stratification (influenced by surface heat fluxes), and non-linear advection significantly influence the circulation.
Along the western boundary the bottom pressure torque plays a key role in steering the flow, and the vorticity input by the bottom pressure torque is partly redistributed by non-linear advection.
Our study highlights the importance of both atmospheric forcing and oceanic dynamical processes in driving long-term gyre circulation responses to the NAO.
\end{abstract}

\section*{Plain Language Summary}
The North Atlantic Oscillation (NAO) is an atmospheric pattern that plays a critical role in setting weather and climate patterns across the North Atlantic region.
The NAO affects the ocean's surface through variations in winds and air-sea heat exchanges, which in turn impact large-scale circulatory flows called gyres.
Gyres transport heat, carbon dioxide, and nutrients within and across ocean basins.
However, the contributions of small-scale swirling flows (eddies) and the seafloor in modulating the gyre circulation to the NAO are not entirely clear.
Here, we analyze global ocean-sea ice simulations to understand long-term changes in the North Atlantic subtropical and subpolar gyres in response to NAO forcing.
We find that both gyres strengthen under positive NAO conditions.
The NAO-induced changes in winds strongly control the gyre circulation in the mid-latitudes ($20^{\circ}-45^{\circ}$N), but are less important in subpolar regions.
Pressure differences on the seafloor, density distribution via thermal wind (and controlled by air-sea heat exchanges), and eddy interactions via non-linear momentum transport are crucial in steering the subpolar gyre circulation.
Our study demonstrates that both atmospheric forcing and oceanic dynamical responses to that forcing are essential in shaping the response of the gyre circulation due to NAO forcing.

\section{Introduction}

The North Atlantic Oscillation (NAO) is a dominant mode of atmospheric variability that influences the North Atlantic Ocean on sub-seasonal to multi-decadal timescales, particularly in boreal winter \cite{Hurrell1995, Hurrell2001}.
The NAO plays a crucial role in regulating weather patterns across Europe and the United States \cite{Hurrell1997, Deser2017}.
A positive NAO phase is characterized by stronger-than-average air pressure difference between Iceland and the Azores; lower-than-average air pressure difference is observed in a negative NAO phase.
These pressure anomalies drive significant changes in the atmospheric forcing at the ocean's surface, including wind patterns and surface heat fluxes \cite{Visbeck2003}.
The changes in surface forcing, in turn, impact the large-scale ocean circulation, including the subtropical and subpolar gyres \cite{Marshall2001-NAO, Eden2001, Lohmann2009, Wolfe2019}.
Therefore, through the gyres the NAO also influences the redistribution of heat \cite{Khatri2022}, nutrients \cite{Oschlies2001}, and carbon dioxide \cite{Keller2012} in the global ocean.
The goal of this paper is to better understand the physical processes by which the NAO impacts the North Atlantic gyre circulation.

Earlier studies have identified several key changes in the North Atlantic circulation resulting from NAO-induced variations in the surface forcing.
\citeA{Curry2001} demonstrated that both the subtropical and subpolar gyres weakened during the negative NAO phase in the late 1960s due to weakened westerlies and intensified during the positive NAO phase in the 1990s.
Through hydrography analysis, \citeA{Curry2001} revealed that both gyres exhibit similar variability (25-33\% change in the circulation strength between the two phases).
While wind stress is thought to play the dominant role in driving variability in the subtropical gyre, the subpolar gyre is thought to also be affected by NAO-induced surface heat flux anomalies \cite{Hakkinen2004, Khatri2022}.
In addition, the Gulf Stream's separation latitude is observed to shift northward during positive NAO phases and southward during negative NAO phases \cite{Chaudhuri2011, Hameed2018, Chi2019, Wolfe2019, Silver2021}.
These shifts are associated with a meridional displacement in the zero wind stress curl latitude \cite{Marshall2001-NAO}.
Thus, the NAO-induced surface forcing anomalies impact both the strength and the spatial structure of the North Atlantic gyre circulation.

The response of the North Atlantic gyres to the NAO occurs across a range of timescales.
Several studies \cite{Curry2001, Marshall2001-NAO, Coetlogon2006} have identified a two-stage gyre response to NAO variability: a rapid ($< 2$~months) barotropic response driven by near-surface Ekman transport, followed by a slower ($2-8$ years) baroclinic adjustment by Rossby waves \cite{Marshall2013}.
The long-term response is complicated by the interplay between these two (i.e., fast Ekman and slow baroclinic) processes. 
One approach to quantifying the long-term response of the gyre to NAO anomalies involves applying synthetic, long-lived NAO forcings in a numerical ocean model. For example, \citeA{Eden2001} and \citeA{Lohmann2009} utilized decadal NAO forcings to demonstrate that apart from the baroclinic adjustment by Rossby waves (a direct atmospheric forcing impact), the delayed response of the gyre to the NAO can also be attributed to an imprint of surface heat fluxes which imparts `memory' to the ocean through the mixed layer. This imprint can cause a non-local impact on the ocean circulation on long timescales (e.g., through heat advection from lower to higher latitudes \cite{Khatri2022}).
The quantification of the long-term NAO-induced subpolar gyre circulation changes at high latitudes is especially important because of the gyre's interconnectedness with the Atlantic Meridional Overturning Circulation (AMOC; e.g., see \citeA{Lohmann2009}, \citeA{Delworth2016}, and \citeA{Yeager2020}).
\citeA{Kim2020} and \citeA{Kim2024} applied surface heat flux anomalies over the subpolar gyre associated with the NAO for a decade, albeit only during the winter period. They found a critical role of the thermocline processes (e.g., changes in stratification) in shaping the gyre circulation, especially in the western boundary region.
However, these studies do not explore the relative role of key dynamical features, such as topography, mesoscale eddies, and viscosity, in shaping the NAO-induced variations in these gyres.
In this paper, we aim to better understand the long-term gyre response to sustained NAO anomalies using a vorticity budget.


Vorticity budgets are derived from the horizontal momentum equations and can reveal dynamical balances between physical processes that govern the large-scale ocean circulation \cite{Waldman2023, Khatri2024}.
Classical theories of gyre circulation used the Sverdrup balance; a simplified form of the vorticity balance between depth-integrated meridional flow in the gyre and the curl of the wind stress \cite{Sverdrup1947}.
Sverdrup balance holds in the gyre's interior \cite{Gray2014}, where the depth-integrated flow interacts negligibly with the bathymetry and largely follows geostrophic balance.
In contrast, the return flow of the gyre is achieved via western boundary currents, where the Sverdrup balance is significantly disrupted and processes such as bathymetric stresses \cite{Stommel1948} and mesoscale eddies \cite{Munk1950} support these boundary currents.
In regions of uneven topography, the Earth can guide currents via bottom pressure anomalies.
This concept was first used to explain the momentum balance of the Antarctic Circumpolar Current by \citeA{Munk1951} but was later found to be crucial in steering gyres through an input of bottom pressure torque, especially in regions of steep topography such as in the western boundary \cite{Holland1967, Holland1973, Hughes2001}.
The bottom pressure torque reflects the influence of topography in steering the flow up/down the slope \cite{Salmon1998}.
With this addition, the large-scale gyre circulation can be described as a `topographic--Sverdrup balance':
\begin{equation}
    \beta V = \frac{\bzhat \bcdot (\bnabla \times \btau_s)}{\rho_0} + \frac{\JpbH}{\rho_0},
    \label{eq:topo-Sverdrup-balance}
\end{equation}
where $\beta$ is the meridional gradient of the Coriolis frequency $f$, $V$ is the depth-integrated meridional transport, $\bzhat$ is the unit vector in the vertical, $\btau_s$ is the horizontal wind stress, $\rho_0$ is the mean seawater density,  $p_b$ is the bottom pressure, and $H$ is the ocean's depth. The term $\JpbH / \rho_0$ is the bottom pressure torque.
For a flat bottom, $\bnabla H = 0$ and the topographic-Sverdrup balance reduces to the classical Sverdrup balance.
We note that the bottom pressure torque should not be misinterpreted as an external driver of gyre circulation.
Instead, it is the surface forcing that externally induces vorticity and thus drives the gyre circulation, which in turn is guided by variations in topography and bottom pressure \cite{Waldman2023}.

The topographic-Sverdrup balance is a good steady-state approximation for the large-scale gyre circulation away from boundary currents \cite{Holland1967}.
However, other processes may dominate the vorticity balance on smaller scales \cite{Sonnewald2019}.
For example, mesoscale eddies can advect vorticity anomalies within the gyre \cite{Holland1978, Rhines1979}.
In models, viscosity also acts to diffuse momentum and dissipate energy as a replacement for unresolved scales of motion \cite{Jochum2008}, and impacts the vorticity budget through the curl of the viscous friction terms.
These processes occur on small scales, and thus the grid scale balance can be quite different to the gyre-scale vorticity balance.
In other words, since these processes vary on small scales, it becomes difficult to interpret their control on the large-scale gyre circulation by analyzing vorticity budgets at each grid cell.
To this end, we analyze dynamically motivated area-integrated vorticity budgets to assess how small-scale processes such as non-linear advection and viscosity influence the large-scale gyre circulation.

Several studies have analyzed complete forms of the vorticity budget at high-resolutions in both global settings \cite{Sonnewald2019, Khatri2024} and with a particular focus on the North Atlantic Ocean \cite{Yeager2015, Schoonover2016, Bras2019, Corre2020}.
One limitation of using canonical forms of vorticity budgets is that they do not explicitly include the gyre strength as a term.
Thus, it becomes difficult to relate the gyre circulation to physical processes such as winds, non-linear advection, and topography.
%
%
In this paper, we rewrite the barotropic vorticity budget (described in section~\ref{section:methods}) to expose a novel relationship between gyre strength and the relevant dynamical processes.
We investigate the relative importance of these processes in shaping the time-integrated gyre response to NAO forcing using an eddy-permitting ocean-sea ice model, as outlined in section~\ref{section:model-setup}.
We first analyze a time-mean version of the modified budget from a control simulation in section~\ref{section:control-bvb-analysis}, followed by a comparative analysis of the time-integrated response of both the subtropical and subpolar gyres to NAO forcing anomalies in section~\ref{section:perturbations-bvb-analysis}.
We summarize our key findings in section~\ref{section:summary}, where we also compare our work with past studies.
Finally, we discuss the broader implications of our analysis for predicting long-term NAO-induced variations in the North Atlantic subtropical and subpolar gyre circulation.


\section{Model setup}
\label{section:model-setup}
We use a global configuration of the Modular Ocean Model~6~(MOM6) \cite{Adcroft2019} at a $0.25^{\circ}$ horizontal resolution, which partially resolves mesoscale eddies in the tropics and mid-latitudes \cite{Hallberg2013}.
The model is discretized into 75 unevenly spaced vertical layers, with layer thicknesses increasing from 2~meters at the surface to 500~meters in the ocean's abyss.
The model uses a $z^{\star}$ vertical coordinate, which is a rescaled height coordinate that treats the time-dependent free surface as a coordinate surface \cite{Adcroft2004}.
Frictional stress at the bottom is modeled using a quadratic drag law with the near-bottom velocities supplemented with a constant velocity of $0.1\,\textrm{m}\,\textrm{s}^{-1}$.
This ocean model is coupled to the Sea Ice Simulator~v2, a dynamic/thermodynamic sea ice model \cite{Adcroft2019}.
We do not parametrize mesoscale eddy transport and stirring.
We use the default OM4p25 configuration of MOM6; details on the model's parameters and choices can be found in the paper by \citeA{Adcroft2019}.

\begin{figure}[ht]
    \includegraphics[width=1.0\textwidth]{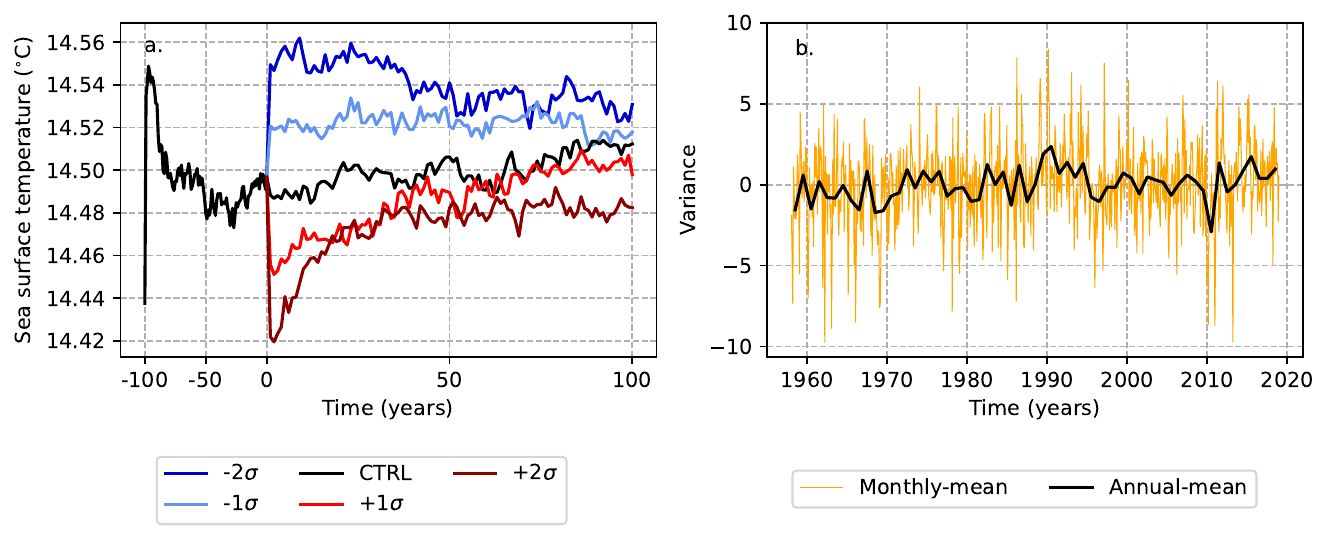}\vspace{-1em}
    \caption{Outline of the NAO-forced perturbations.
    (a)~Time series of the sea surface temperature obtained by spatially averaging in the North Atlantic Ocean, illustrating the model spinup. Simulation time is referenced with respect to the beginning of the perturbation experiments.
    (b)~NAO-index derived from the first principal component of the monthly-mean (yellow) and yearly-mean (black) sea level pressure anomalies in the North Atlantic Ocean. The yearly-mean index is scaled to unit variance.
    }
    \label{fig:expt_setup}
\end{figure}

\begin{figure}
\includegraphics[width=1.0\textwidth]{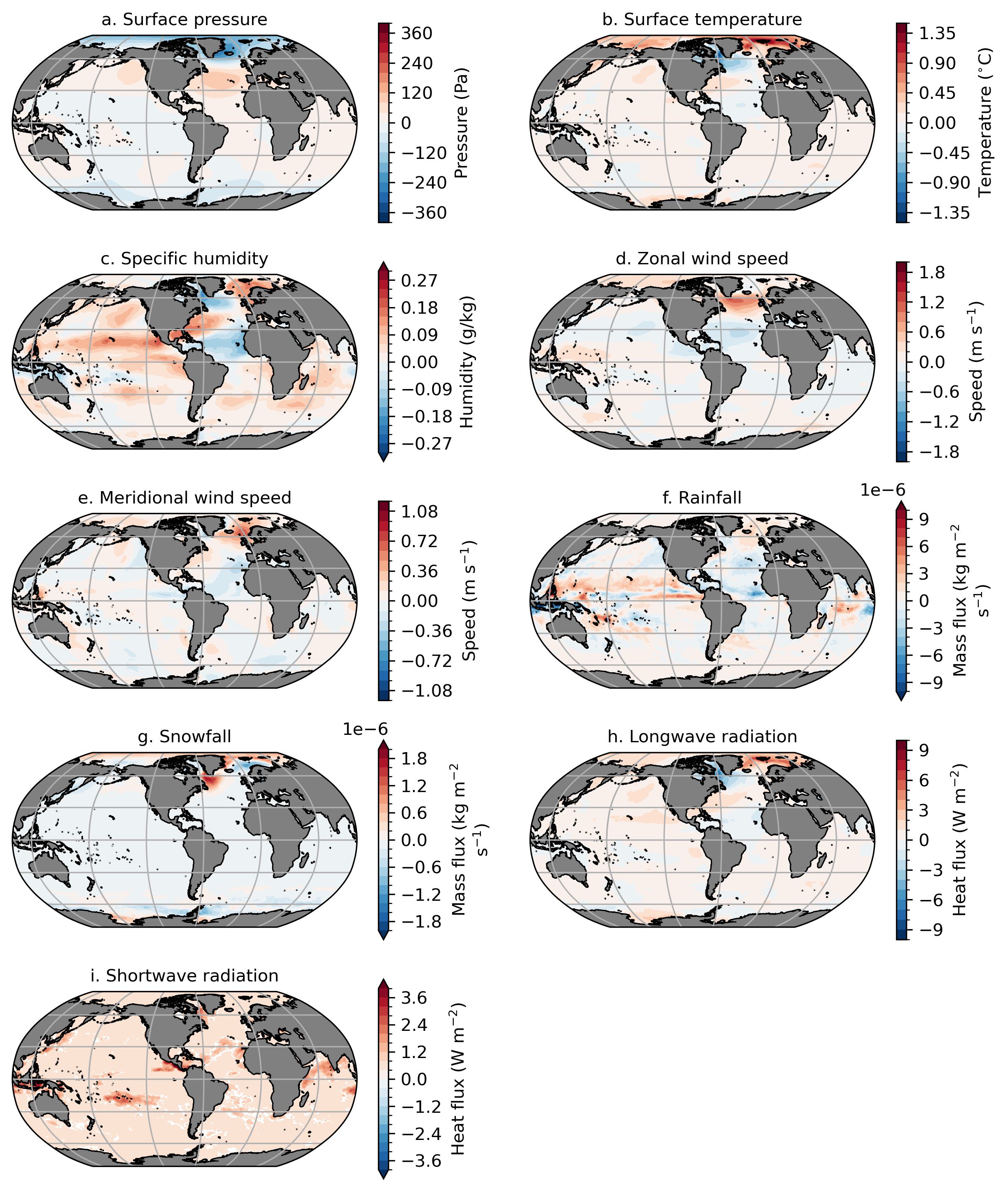}
    \caption{Regression maps of atmospheric forcing field anomalies representing changes during a $+1\sigma$ NAO state.
    (a)~Surface pressure (Pa),
    (b)~Surface air temperature ($^{\circ}$C),
    (c)~Specific humidity (kg/kg),
    (d)~Zonal wind speed (m s$^{-1}$),
    (e)~Meridional wind speed (m s$^{-1}$),
    (f)~Rainfall (kg m$^{-2}$ s$^{-1}$),
    (g)~Snowfall (kg m$^{-2}$ s$^{-1}$),
    (h)~Longwave radiation (W m$^{-2}$), and
    (i)~Shortwave radiation (W m$^{-2}$).
    Red (positive) colors in panels~(f-g) indicate mass entering the ocean, while blue (negative) colors indicate mass being removed from the ocean. Similarly, the red colors in panels~(h-i) indicate heat entering the ocean, while the blue colors indicate heat loss at the ocean's surface.
    }
    \label{fig:NAO-surface-forcing}
\end{figure}

We run a series of global ocean--sea ice simulations driven by a repeat-year forcing (at 3-hourly temporal resolution) based on the JRA55-do atmospheric reanalysis dataset \cite{Tsujino2018}.
We use May 1990 to April 1991 as our repeat year as it is one of the most neutral years in the JRA55-do dataset in terms of major climate modes of variability \cite{Stewart2020}.
The control simulation is run for 101 years (black line in Fig.~\ref{fig:expt_setup}a), after which we branch off four experiments with temporally-invariant forcing perturbations (i.e., step function perturbations) resembling specific NAO states.
To create these perturbations, we use the standardized approach in the literature (e.g., see \citeA{Hurrell1995} and \citeA{Wanner2001}) by defining an NAO index as the leading principal component of the JRA55-do sea level pressure anomalies in the North Atlantic basin (100$^\circ$W--10$^\circ$E, 10$^\circ$N--75$^\circ$N).
Most past studies focus on winter sea level pressure anomalies to calculate the NAO index, as it is most clearly observed during that period.
However, since we are motivated to analyze a time-integrated gyre response to the NAO, we consider the first principal component of the de-seasonalized sea level pressure anomalies for the whole year.
The resulting NAO index (black line in Fig.~\ref{fig:expt_setup}b) explains about 38\% variance, which is marginally less than than if we would have taken only the winter period (41\% variance; not shown).
For comparison, the next two EOF modes obtained from the yearly-mean sea-level pressure anomalies respectively explain 20\% and 13\% of the variance.
Therefore, our NAO index is distinct in capturing a major mode of variability \cite{Dommenget-Latif-2002}.
Finally, we note that the global warming signal can also be removed from the NAO index by detrending the sea level pressure anomalies during the $1958-2018$ period.
We have not done any detrending here, as the influence on the NAO index variance is less than $0.3\%$.

We then use our NAO index to create perturbed surface forcing variables.
We normalize the NAO index to unit variance and linearly regress it against each JRA55-do atmospheric forcing variable (except mass flux due to river runoff and frazil formation, which are small and hence are not modified) between years 1958-2018.
The regression maps in Fig.~\ref{fig:NAO-surface-forcing} indicate anomalies in the surface forcing variables during a $+1\sigma$ NAO index.
The largest regressions for each variable are generally found in the North Atlantic Ocean.
However, some forcing variables display noticeable remote variability, e.g., a positive NAO phase is associated with Arctic warming (Fig.~\ref{fig:NAO-surface-forcing}b) and increased rainfall in the tropical Pacific Ocean (Fig.~\ref{fig:NAO-surface-forcing}f), consistent with well-recognized teleconnections of the NAO with the Arctic Oscillation \cite{Thompson1998} and the El Ni\~no Southern Oscillation \cite{King2023}.

The regression maps in Fig.~\ref{fig:NAO-surface-forcing} form the cornerstone of our experimental setup.
These perturbations are either added or subtracted from the CTRL's surface forcing variables to mimic a positive or negative NAO index respectively.
For instance, the simulation labeled as $-2\sigma$ (dark blue line in Fig.~\ref{fig:expt_setup}a) is generated by doubling the anomaly maps in Fig.~\ref{fig:NAO-surface-forcing} and subtracting them from the control's surface forcing variables.
Our experiments, coupled with the barotropic vorticity budget described below, allow a clear understanding of the mechanisms that dominate the gyres' equilibrium response, which cannot be analyzed in an interannually-forced simulation.

\section{Methods}
\label{section:methods}
\subsection{Gyre strength}
\label{subsection:gyre-strength}
\begin{figure}[ht]
    \includegraphics[width=1.0\textwidth]{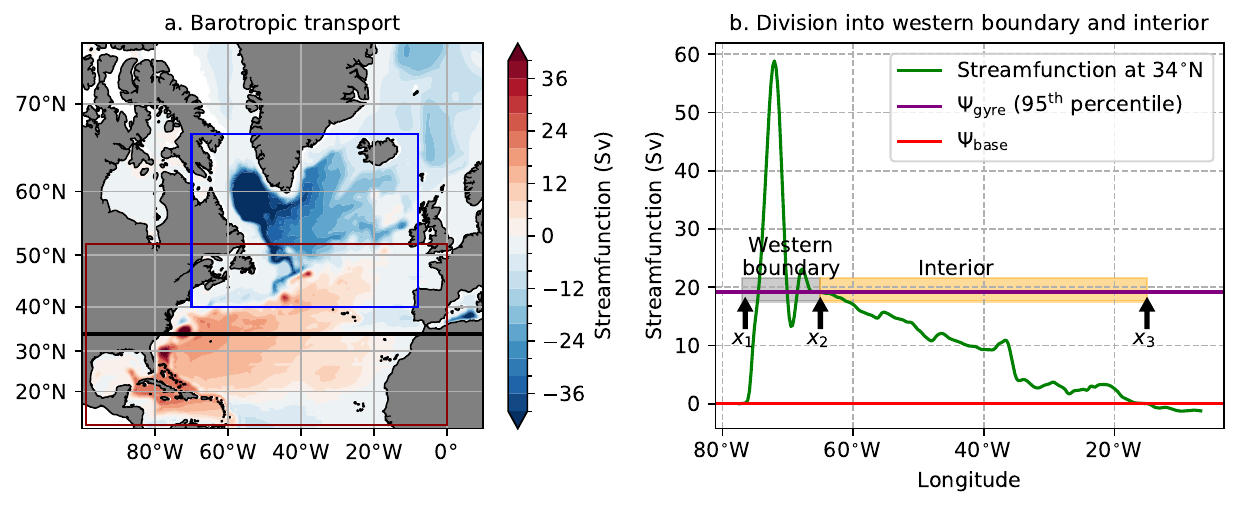}\vspace{-1em}
    \centering
    \caption{(a).~Barotropic streamfunction for the North Atlantic Ocean. Red and blue boxes denote regions over which the 95$^{\rm{th}}$ percentile is computed to obtain subtropical and subpolar gyre strength metrics respectively. Black line indicates 34$^{\circ}\rm{N}$ for which we show the gyre's division as an example.
    (b).~Division of the subtropical gyre at 34$^{\circ}\rm{N}$ into a western boundary (gray shading) and interior flow (orange shading). This division is based on the easternmost occurrence of $\Psi_{\rm{gyre}}$ at each latitude.}
    \label{fig:gyre-methods}
\end{figure}

We begin by defining a gyre strength metric via a barotropic streamfunction $\Psi$ -- that is the cumulative integral of the vertically-integrated meridional volume transport from the western to eastern ends of the North Atlantic basin (Fig.~\ref{fig:gyre-methods}a).
Similar to the study by \citeA{Bhagtani2023}, we select the 95$^{\textrm{th}}$ percentile (based on area) of the streamfunction for any given experiment to derive a singular metric for gyre strength; we denote this as $\Psi_{\textrm{gyre}}$.
In the following subsections, we show that through appropriate choices of integration area, $\Psi_{\textrm{gyre}}$ can be directly and quantitatively related to the sum of the terms in the barotropic vorticity balance.

\subsection{Barotropic vorticity budget}
Following \citeA{Khatri2024}, we define the barotropic vorticity budget as the curl of the Bousinessq-hydrostatic depth-integrated horizontal momentum equation:
\begin{multline}
    \underbrace{\;\beta V\;\vphantom{\frac{\bzhat}{\rho_0}}}_{\substack{\textrm{Planetary}\\\textrm{vorticity}\\\textrm{advection}}} = \underbrace{\frac{\JpbH}{\rho_0}}_{\substack{\textrm{Bottom pressure}\\\textrm{torque}}} + \underbrace{\frac{\bzhat \bcdot(\bnabla \times \btau_s)}{\rho_0}}_{\textrm{Wind stress curl}} \underbrace{- \frac{\bzhat \bcdot(\bnabla \times \btau_b)}{\rho_0}}_{\textrm{Bottom drag curl}} + \underbrace{\bzhat \bcdot (\bnabla \times \badv)\vphantom{\frac{\bzhat}{\rho_0}}}_{\textrm{Non-linear advection}} \\ + \underbrace{\bzhat \bcdot (\bnabla \times \bvisc)\vphantom{\left(\frac{\partial \bcalU}{\partial t}\right)}}_{\textrm{Horizontal friction}}
    \underbrace{- \bzhat \bcdot \left(\bnabla \times \frac{\partial \bcalU}{\partial t}\right)}_{\textrm{Vorticity tendency}} \underbrace{- f \frac{Q_m}{\rho_0}}_{\textrm{Mass flux}} + \underbrace{f \frac{\partial \eta}{\partial t}\vphantom{\left(\frac{\partial \bcalU}{\partial t}\right)}}_{\substack{\textrm{Surface}\\\textrm{elevation}\\\textrm{tendency}}},
    \label{eq:BVB}
\end{multline}

\noindent where $\btau_s$ is the surface stress on the ocean by the atmosphere and sea ice, $\btau_b$ is the bottom stress, $\badv$ is the vertically-integrated non-linear advection term, $\bvisc$ the vertically-integrated viscosity term, $Q_m$ is the mass flux at the ocean's surface, $\bcalU = U\bxhat + V \byhat $ is the vertically-integrated horizontal flow, and $\eta$ is the surface elevation above the mean sea level.
We ensure that the budget closes at each grid cell (where all terms are defined) down to machine precision error by making use of the terms leading to a closed momentum budget at each grid point (see \citeA{Khatri2024} for details).

Other forms of the vorticity budget can also be constructed, depending on whether depth-averaging or integration is used, or what order computations are performed in.
Each construction has its own theoretical and numerical pros and cons, and are carefully reviewed by \citeA{Waldman2023}.
Our barotropic vorticity budget is derived by first integrating in depth, followed by taking a curl.
This order of operation ensures that we account for all grid cells near rough topography.
However, residuals may be present in individual terms.
For example, \citeA{Khatri2024} found large numerical discretization errors in the planetary vorticity advection term (associated with the curl operation involved in the term's calculation) that were at least two orders of magnitude larger than the mass flux and tendency terms combined.
These errors are almost entirely offset by opposing large numerical errors in the bottom pressure torque term.
\citeA{Styles2022} discuss several approaches to reduce the numerical errors in the bottom pressure torque, including the use of an Arakawa B-grid or a terrain-following vertical coordinate.

The right side of the barotropic vorticity budget highlights the interplay between all physical processes that affect an arbitrary depth-integrated meridional flow at each grid cell.
Area-integration of this budget thus yields the large-scale balance affecting the net meridional flow within that area.
Different choices of area-integration exist (e.g., see \citeA{Stewart2021} for a comparison between depth, streamfunction, and sea surface height as parameters for integration).
We choose streamfunction contours (that are approximately stationary in time) averaged over the last 30 years as our area of integration, which helps us include gyre strength as an explicit term in the barotropic vorticity budget by rewriting the $\beta V$ term.
Then, the meridional flow $V = \partial_x \Psi$ (where $\Psi$ is the barotropic streamfunction defined in section~\ref{subsection:gyre-strength}) for any given latitude can be integrated as:
\begin{equation}
    \int_{x_{\min}}^{x_{\max}} V \,\mathrm{d}x = \Psi \Big\vert_{x_{\max}} - \Psi \Big\vert_{x_{\min}},
    \label{eq:betaV-streamfunction-2}
\end{equation}
noting that $\beta$ does not vary with longitude ($\partial_x\beta=0$) and $x_{\rm{min}}$ and $x_{\rm{max}}$ define the longitudes of integration.
Equation~\eqref{eq:betaV-streamfunction-2} shows that simply integrating $\beta V$ over the streamfunction contour that encloses the gyre yields zero (since this choice implies $\Psi \big\vert_{x_{\max}}$ = $\Psi \big\vert_{x_{\min}}$ in Eq.~\eqref{eq:betaV-streamfunction-2}), and thus is not very useful in linking the gyre strength to various dynamical processes.

Instead, motivated by past studies that highlight different dynamical balances in the western boundary and in the gyre's interior in the North Atlantic \cite{Sonnewald2019, Bras2019, Corre2020}, we choose to separately analyze the area-integrated vorticity balance for these two above-mentioned regions.
We define the outer boundary of the gyre using an arbitrary closed streamfunction contour, $\Psi_{\rm{base}}$.
We further divide the gyre into a western boundary and interior region using $\Psi_{\rm{gyre}}$ (Fig.~\ref{fig:gyre-methods}b).
For each latitude, the gyre's western boundary (gray horizontal bar in Fig.~\ref{fig:gyre-methods}b) is bounded by $x_{\min} = x_{\rm{1}}$ and $x_{\max} = x_{\rm{2}}$ such that $\Psi \big\vert_{x_{\max}} = \Psi_{\rm{gyre}}$ and $\Psi \big\vert_{x_{\min}} = \Psi_{\rm{base}}$.
For cases where $\Psi \big\vert_{x_{\max}} = \Psi_{\rm{gyre}}$ is true for multiple longitudes (as shown in Fig.~\ref{fig:gyre-methods}b), we pick the easternmost longitude that satisfies the condition (i.e., we choose to include the area where the streamfunction exceeds the 95$^{\rm th}$ percentile within the western boundary region rather than the interior).
Then, the gyre's interior (orange horizontal bar in Fig.~\ref{fig:gyre-methods}b) at each latitude is bounded by $x_{\min} = x_{\rm{2}}$ and $x_{\max} = x_{\rm{3}}$ such that $\Psi \big\vert_{x_{\max}} = \Psi_{\rm{base}}$ and $\Psi \big\vert_{x_{\min}} = \Psi_{\rm{gyre}}$.
We note that the western boundary region only exists for latitudes where $\vert \Psi \vert > \vert \Psi_{\rm{gyre}} \vert$ for any longitude within the gyre's boundary.

The above approach ensures that the area-integrated $\beta V$ term is non-zero while being equal and opposite for the two regions. From Eq.~\eqref{eq:betaV-streamfunction-2}, changes in $\Psi_{\mathrm{gyre}}$ (and therefore, the barotropic vorticity budget) can only occur if there is a net flow between the two regions. In other words, $\Psi_{\mathrm{gyre}}$ cannot change due to re-circulations that exist entirely within one region. Using this approach, we can rewrite the barotropic vorticity budget (Eq.~\eqref{eq:BVB}) to determine the relative contributions of each term on the gyre strength $\Psi_{\rm gyre}$ for the western boundary and interior regions:

\begin{align}
    \Psi_{\rm gyre} - \Psi_{\textrm{base}} &= -\left( \int_{A_{\rm WBC}} \mathcal{F} \, \mathrm{d}A \right)  \left({\int_{y_{\min}, \rm WBC}^{y_{\max}, \rm WBC} \beta \, \mathrm{d}y} \right)^{-1}  \nonumber\\
    &= \left({\int_{A_{\rm interior}} \mathcal{F} \, \mathrm{d}A} \right) \left({\int_{y_{\min}, \rm interior}^{y_{\max}, \rm interior} \beta \, \mathrm{d}y} \right)^{-1},
\label{eq:gyre-strength-bvb}
\end{align}

\noindent where $\mathcal{F}$ collectively denotes all the right hand side terms in Eq.~\eqref{eq:BVB}.
Equation~\eqref{eq:gyre-strength-bvb} quantifies the contribution of each term in the barotropic vorticity budget to the gyre strength.

It is important to discuss one constraint on the closure of the rewritten barotropic vorticity budget.
While some model outputs relevant to estimating the barotropic vorticity budget (Eq.~\eqref{eq:BVB}) are natively present on the tracer grid, others are present on the momentum grid.
This causes a few terms on the right hand side of the barotropic vorticity budget (Eq.~\eqref{eq:BVB}) that involve horizontal gradients (e.g., bottom pressure torque) to be undefined for a small fraction of grid cells adjacent to land masses (for the lateral resolution we use here this fraction is at most 4\%).
To ensure closure at each grid cell, \citeA{Khatri2024} removed these grid cells from the analysis.
However, meridional flow exists in these grid cells.
The planetary vorticity advection term in Eq.~\eqref{eq:gyre-strength-bvb} is formulated using the monthly-mean barotropic streamfunction output from the model, which is defined on all ocean grid cells (including those where the budget is not defined).
The right hand side of Eq.~\eqref{eq:gyre-strength-bvb} is unable to represent this flow (since they are direct outputs from the barotropic vorticity budget), and thus this flow is referred to as residual flow as it appears as a residual in the budget.

We examine the relative importance of the terms in Eq.~\eqref{eq:gyre-strength-bvb} for the control simulation in the next section, followed by a comparative analysis of the budget across different NAO states.

\section{Results: Vorticity budget in the control simulation}
\label{section:control-bvb-analysis}
In this section, we study the CTRL simulation's time-mean vorticity budget using Eq.~\eqref{eq:gyre-strength-bvb} for the North Atlantic subtropical and subpolar gyres for years $71-100$; see Fig.~\ref{fig:expt_setup}).

\subsection{Subtropical gyre}

\begin{figure}[h!]
    \includegraphics[width=1.0\textwidth, trim = {17cm 4cm 17cm 1.5cm}, clip]{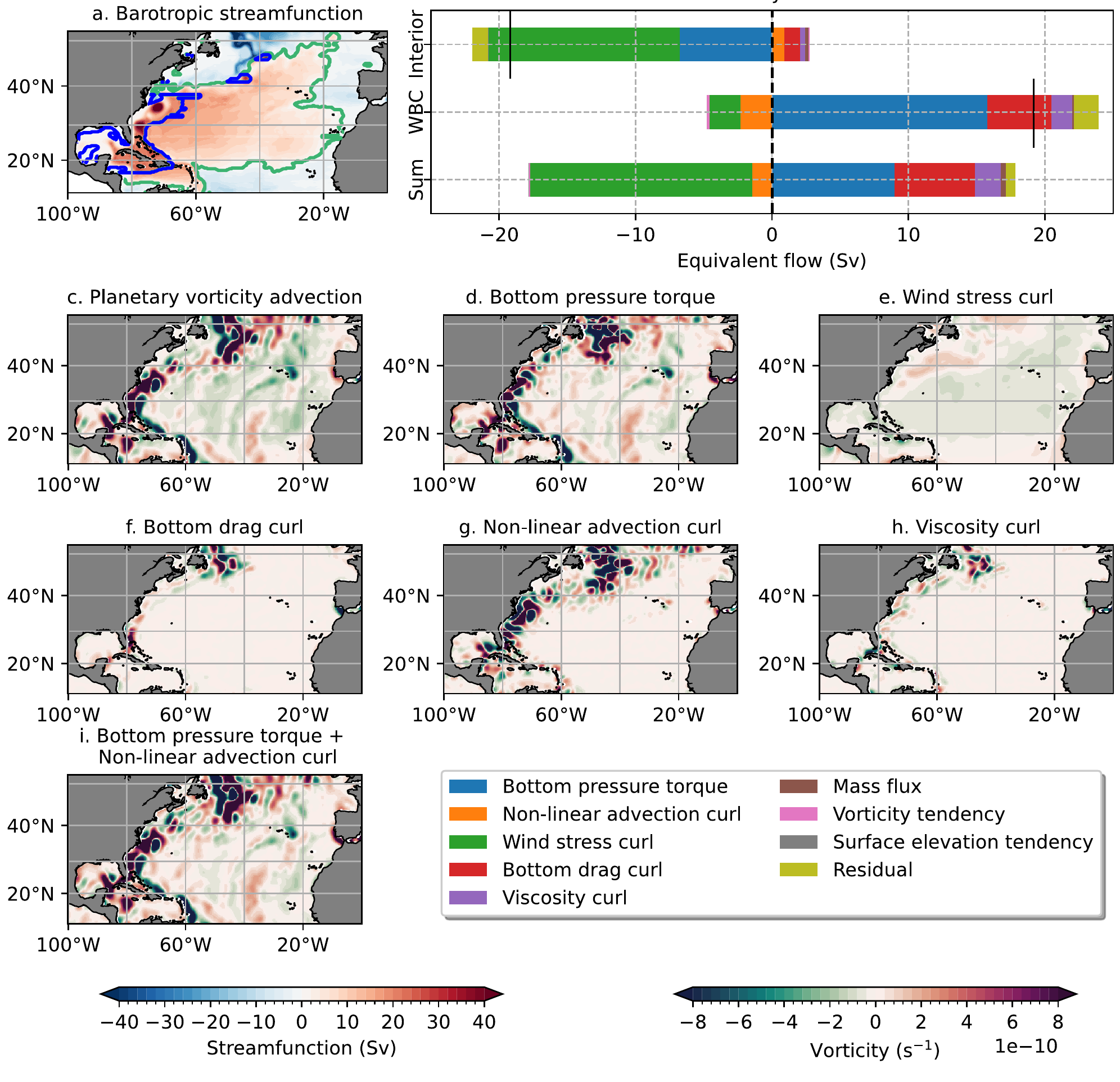}\vspace{-1em}
    \caption{(a).~Barotropic streamfunction for the North Atlantic subtropical gyre (bottom left colorbar). Blue and green contours highlight the western boundary and interior regions respectively.
    (b).~Equivalent flow induced by each term on the right hand side of the barotropic vorticity budget (Eq.~\eqref{eq:gyre-strength-bvb}) for the western boundary and interior regions, along with their sum.
    The solid black line denotes $|\Psi_{\textrm{gyre}} - \Psi_{\textrm{base}}| = 19.18$~Sv.
    (c-h).~Vorticity maps (filtered using a Gaussian kernel of 2$^{\circ}$~radius) for key terms in the barotropic vorticity budget (bottom right colorbar).
    (i)~Sum of bottom pressure torque and non-linear advection curl highlighting that the residual of these two terms balance the planetary vorticity advection term (panel~c) in the western boundary.}
    \label{fig:ctrl-bvb-subtropics}
\end{figure}

The 95$^{\text{th}}$ percentile of the streamfunction, our metric for the gyre strength, is $19.18$~Sv for the subtropical gyre in the control simulation.
We consider $\Psi_{\rm{base}} = 0$~Sv as the bounding area of integration.
As described in the previous section, we divide the subtropical gyre into a western boundary and interior region, as shown respectively by the blue and green contours in Fig.~\ref{fig:ctrl-bvb-subtropics}a.
The interior contains a broad equatorward flow, which is balanced by a narrow northward flowing western boundary current.
For each region, terms on the right hand side of Eq.~\eqref{eq:gyre-strength-bvb} that have the same sign as $\beta V$ strengthen the gyre flow, while terms that have the opposite sign weaken the gyre flow.
We note that the force balances in both the western boundary region and in the gyre's interior is robust to small changes in $\Psi_{\rm{base}} (\mathcal{O}(0.1\,\mathrm{Sv}))$.
For large changes in $\Psi_{\rm{base}}$, the relative contribution of wind stress and bottom pressure torque changes in the interior, and is outlined later in this section.

A topographic-Sverdrup balance exists in the gyre's interior (Fig.~\ref{fig:ctrl-bvb-subtropics}b; also see a rough balance between panels~(c),~(d) and~(e)), consistent with past literature \cite{Sonnewald2019, Khatri2024}.
The wind stress curl drives 73\% of the equatorward flow through an anticyclonic vorticity input equivalent to $14.0$~Sv via Ekman convergence \cite{Ekman1905} (green bar in Fig.~\ref{fig:ctrl-bvb-subtropics}b).
This vorticity input quantifies the circulation strength if the gyre's interior flow were completely Sverdrupian.
The spatially uniform vorticity input by the wind stress curl (Fig.~\ref{fig:ctrl-bvb-subtropics}e) means that this input is sensitive to the chosen value of $\Psi_{\rm{base}}$: picking a larger value of $\Psi_{\rm{base}}$ skews the gyre's bounding streamfunction towards the west and the circulation becomes less Sverdrupian.
Under such circumstances, the effect of topography through bottom pressure anomalies becomes more prominent.

With $\Psi_{\rm{base}} = 0\,\mathrm{Sv}$, the bottom pressure torque is the second largest contributor to the vorticity budget ($6.8$~Sv), and is especially noticeable in regions of rough topography such as the mid-Atlantic ridge (Fig.~\ref{fig:ctrl-bvb-subtropics}d).
The bottom pressure torque quantifies the impact of topographic steering that deviates the flow from $f/H$ contours.
More specifically, it quantifies the flow up or down the slope.
In terms of the vertical flow, an anticyclonic vorticity input by the bottom pressure torque, as seen in the gyre's interior, signifies that the net flow descends down the topography, that is, the net vertical flow in the gyre's interior is negative.
Unlike wind stress, the bottom pressure torque exhibits large spatial variability owing to the flow's interaction with small-scale topographic gradients (compare panels~(d) and~(e) in Fig.~\ref{fig:ctrl-bvb-subtropics}).
The bottom drag curl is negligible everywhere in the interior (Fig.~\ref{fig:ctrl-bvb-subtropics}f).
Non-linear advection and viscosity act to respectively redistribute and dissipate vorticity within the system.
The net vorticity input due to both these processes is relatively small in the gyre's interior (Figs.~\ref{fig:ctrl-bvb-subtropics}g-h).
Similarly, the surface mass flux and the tendency terms associated with vorticity and surface elevation input negligible vorticity (Fig.~\ref{fig:ctrl-bvb-subtropics}b) both locally and in an area-averaged sense.
Finally, $1.1$~Sv residual flow exists because our budget (Eq.~\eqref{eq:gyre-strength-bvb}) is not closed for grid cells adjacent to land masses.

Focusing on the western boundary flow, the equivalent flow attributed to the bottom pressure torque term is significant, suggesting a large net flow up the slope (Figs.~\ref{fig:ctrl-bvb-subtropics}b and~d; also see \citeA{Schoonover2016} and \citeA{Bras2019}).
Steep zonal bathymetric slopes under the Gulf Stream lead to strong steering by the bottom pressure through an anticyclonic vorticity input.
The local wind stress curl acts to slow down the Gulf Stream  by $2.3$~Sv (green bar in Fig.~\ref{fig:ctrl-bvb-subtropics}b) by providing small anticyclonic vorticity (Fig.~\ref{fig:ctrl-bvb-subtropics}e).
The next largest share of the circulation is carried by the bottom drag curl, which aligns with previous literature \cite{Hughes2001} suggesting a combined role of the frictional and pressure torques induced by bathymetry in steering the western boundary current.
At first glance, it may seem counter-intuitive that the bottom drag curl supports the western boundary flow, since drag usually removes kinetic energy from the system \cite{Arbic2009}.
This support emerges from the bottom drag curl opposing the large-scale anticyclonic vorticity input by wind stress in the interior, also shown by \citeA{Stewart2021}.

The curl of non-linear advection and viscosity terms display high spatial variability in the Gulf Stream (Fig.~\ref{fig:ctrl-bvb-subtropics}g-h).
While the contribution of the non-linear advection is very small when we integrate its vorticity input over the entire western boundary region (orange bar in Fig.~\ref{fig:ctrl-bvb-subtropics}b), at small-scales it counteracts the vorticity input by the bottom pressure torque \cite{Wang2017}.
The sum of these two terms bears strong spatial resemblance with the planetary vorticity advection term (i.e., meridional flow; Fig.~\ref{fig:ctrl-bvb-subtropics}i), consistent with past studies \cite{Yeager2015, Khatri2024}, and likely results from small-scale meridional flows (captured by the bottom pressure torque) that are non-linearly advected in the ocean \cite{Stewart2021}.

The residual flow ($1.8$~Sv) in the western boundary is larger than in the interior ($1.1$~Sv) owing to a larger number of grid cells near land-sea boundaries where not all terms are defined and are hence omitted from analysis.
Finally, the vorticity input due to surface mass flux and surface elevation tendency terms is negligible almost everywhere within the western boundary region. While the area-averaged vorticity tendency term is negligible, the term can distort the gyre's bounding streamfunction contour through large local values (Fig.~S1). Regions of high eddy activity are especially prone to local distortions in streamfunction, since the vorticity tendency term captures the transient effect of swirling features in the ocean.

We summarize the subtropical gyre circulation as being driven primarily by wind stress curl and modulated by topography through the bottom pressure torque and the bottom drag curl (see the vorticity balance for the entire gyre in Fig.~\ref{fig:ctrl-bvb-subtropics}b; also consistent with \citeA{Schoonover2016} and \citeA{Sonnewald2019}).
The interior circulation follows a topographic-Sverdrup balance, however, in the western boundary region, processes such as the bottom drag curl and the non-linear advection curl terms also influence the gyre circulation.

\subsection{Subpolar gyre}
\begin{figure}[h!]
    \includegraphics[width=1.0\textwidth]{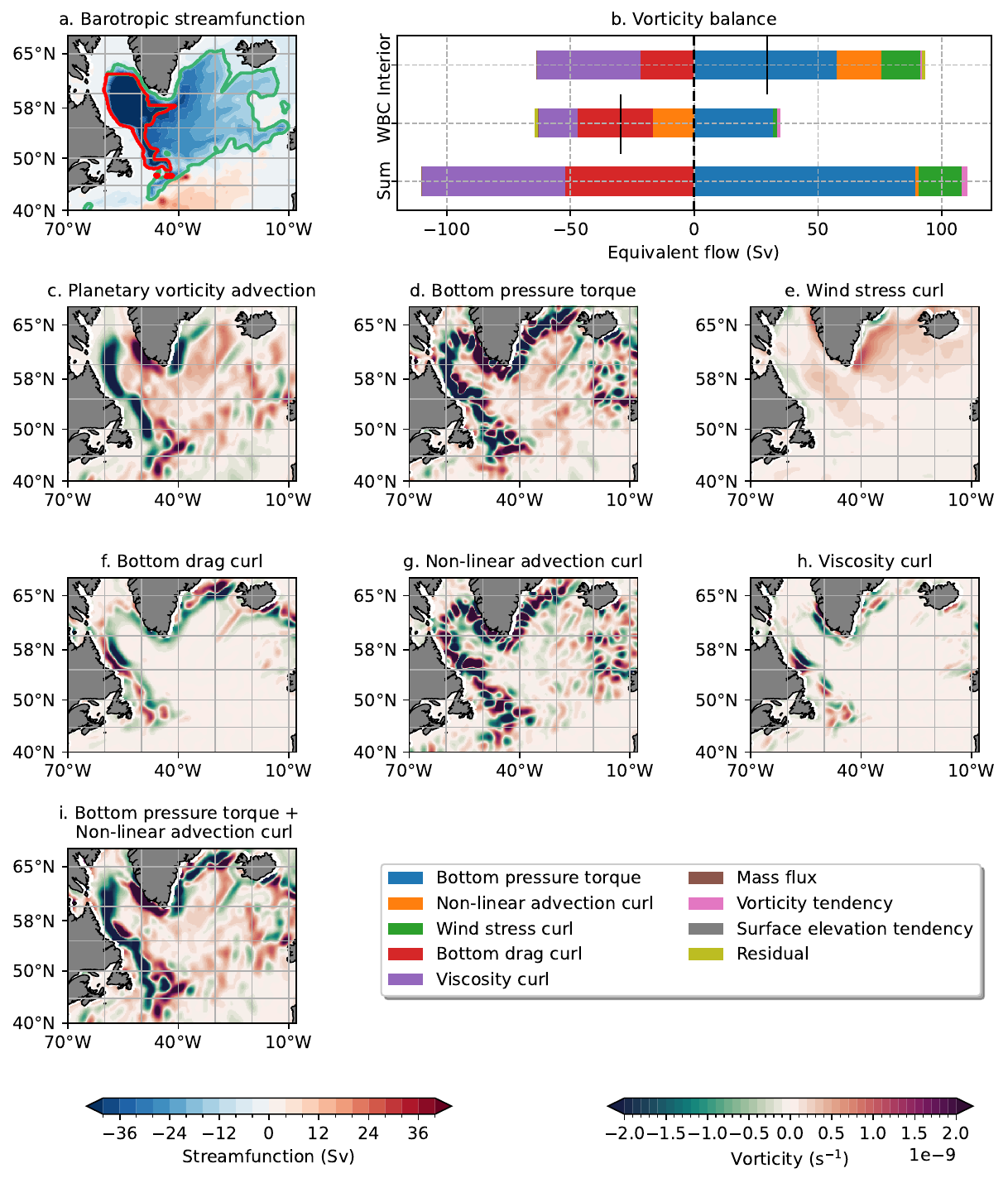}\vspace{-1em}
    \caption{(a).~Barotropic streamfunction for the North Atlantic subpolar gyre (bottom left colorbar). Red and green contours highlight the western boundary and interior regions respectively.
    (b).~Equivalent flow induced by each term on the right hand side of the barotropic vorticity budget (Eq.~\eqref{eq:gyre-strength-bvb}) for the western boundary and interior regions, along with their sum. The dashed black line denotes $|\Psi_{\textrm{gyre}} - \Psi_{\textrm{base}}| = 29.6$~Sv.
    (c-h).~Vorticity maps (filtered using a Gaussian kernel of $2^{\circ}$ radius) for key terms in the barotropic vorticity budget (bottom right colorbar).
    (i)~Sum of bottom pressure torque and non-linear advection curl, highlighting that the residual of these two terms do not balance the planetary vorticity advection term (panel~c) in the western boundary.}
    \label{fig:ctrl-bvb-subpolar}
\end{figure}

The strength of the subpolar gyre, as quantified by the 95$^{\textrm{th}}$ percentile of the streamfunction, is $-39.6~$Sv in the control simulation.
We choose $\Psi_{\textrm{base}} = -10$ Sv because it appropriately constrains our area of integration within the subpolar region; picking a smaller absolute value causes the bounding contour to penetrate into the Arctic.
Similar to the subtropical gyre, we divide the subpolar gyre into a western boundary and interior flow, as shown respectively by the red and green contours in Fig.~\ref{fig:ctrl-bvb-subpolar}a.
For each region, terms on the right hand side of Eq.~\eqref{eq:gyre-strength-bvb} that have the same sign as the meridional flow strengthen the gyre, while terms that have the opposite sign weaken the gyre.
Finally, while our results are robust to small changes ($\mathcal{O}(0.1\,\mathrm{Sv})$) in $\Psi_{\textrm{base}}$, picking a significantly higher negative value of $\Psi_{\textrm{base}}$ (e.g., $-15\,\mathrm{Sv}$) affects the relative contribution of wind stress, bottom pressure torque, and bottom drag curl as outlined in this section.

The flow in the interior subpolar gyre experiences strong deviations from $f/H$ contours (equivalent flow by the bottom pressure torque is $57.4$~Sv) due to significant variations in topography as well as a stronger barotropic nature of the flow (compared to the subtropical gyre) due to weak stratification \cite{Corre2020} (Fig.~\ref{fig:ctrl-bvb-subpolar}b).
In fact, the equivalent flow attributed to the vorticity input by the bottom pressure torque is greater than $|\Psi_{\textrm{gyre}} - \Psi_{\textrm{base}}| = 29.6$~Sv (represented by black vertical lines for the interior and western boundary region in Fig.~\ref{fig:ctrl-bvb-subpolar}b).
The wind stress curl drives a smaller fraction of the gyre circulation ($15.7$~Sv; Fig.~\ref{fig:ctrl-bvb-subpolar}b).
Moreover, the vorticity contribution by wind stress reduces at the expense of the bottom pressure torque when we choose more negative values of $\Psi_{\textrm{base}}$, and other terms become more prominent.
Unlike the subtropical gyre, the subpolar gyre's interior deviates from the topographic-Sverdrup balance.
This deviation is consistent with the vorticity analysis by \citeA{Corre2020} who used a $2~\textrm{km}$ horizontal resolution simulation.
For instance, the curl of non-linear advection term guides a significant portion of the gyre circulation ($18.2$~Sv).
The non-linear advection curl is noticeable almost everywhere in the interior (Fig.~\ref{fig:ctrl-bvb-subpolar}g), suggesting a strong vorticity redistribution by mesoscale eddies \cite{Brandt2004}.
We find large vorticity anomalies in the bottom pressure torque and the subsequent advective redistribution near the Rockall trough (15$^{\circ}$~W, 55$^{\circ}$~N; see Figs.~\ref{fig:ctrl-bvb-subpolar}d and~\ref{fig:ctrl-bvb-subpolar}g).
This region is marked by substantial variations in topography, which promotes large vertical flows that deviate from $f/H$ contours.

Special attention must be paid to the curl of viscosity term, which dissipates vorticity equivalent to $42.0~$Sv of gyre circulation.
In our model, this term is especially large near steep topography and thin currents such as in the northwest corner region in the Grand Banks of Newfoundland and in the Rockall Trough (Fig.~\ref{fig:ctrl-bvb-subpolar}h; also see \citeA{Fischer2018}).
The viscous torque is known to be strongly dependent on the model resolution, numerics and length-scale of interest.
For instance, \citeA{Corre2020} observed a very small contribution of viscosity in altering the overall vorticity budget in their $2$~km horizontal resolution regional model of the North Atlantic, while it was noticeable in the 0.1$^{\circ}$ resolution model by \citeA{Yeager2015}, especially near continental shelves.

The western boundary region (red contour in Fig.~\ref{fig:ctrl-bvb-subpolar}a) captures the Labrador Sea and part of the Irminger Sea, both of which are sites of strong buoyancy loss and dense water formation \cite{Johnson2019}.
This buoyancy loss connects near-surface flows to the abyss through convection, which promotes deep mixed layers and consequently, increased flow-topography interactions in a region of complex topography.
The wind stress curl contributes negligibly ($1.63$~Sv) to the subpolar gyre western boundary region, as expected.
The bottom pressure torque inputs cyclonic vorticity in the western boundary region (equivalent flow $= 31.7\,\mathrm{Sv}$), which indicates that the net vertical flow is up the continental slope.
As a result, the vorticity input by the bottom pressure torque is seen to reduce the western boundary flow.
The bottom drag curl supports the flow by providing cyclonic vorticity input equivalent to $30.4$~Sv.
Choosing a more negative value of $\Psi_{\mathrm{base}}$ causes the continental slope to lie outside of the western boundary region, which reduces the control of bottom pressure torque on the circulation. This reduction is offset by an amplified role of the bottom drag curl. 

The non-linear advection inputs cyclonic vorticity equivalent to $16.8$~Sv in the western boundary flow region.
This vorticity input is similar to the non-linear advective input of $18.2$~Sv in the interior, suggesting that mesoscale eddies could be redistributing vorticity between the western and eastern regions of the gyre \cite{Wang2017}.
There is also a large cancellation of small-scale features between the bottom pressure torque and the non-linear advection term in the western boundary region (Fig.~\ref{fig:ctrl-bvb-subpolar}i).
Unlike in the subtropical gyre, this cancellation does not balance the planetary vorticity advection term, as the bottom drag and viscosity terms are also important (Fig.~\ref{fig:ctrl-bvb-subpolar}f-h).
Both the bottom drag and viscosity curl oppose the flow near the continental shelves (which lies outside our area of integration) and steer the strong western boundary current (Figs.~\ref{fig:ctrl-bvb-subpolar}f-g).
According to \citeA{Yeager2015}, this dipole structure in the viscosity curl broadens the boundary currents by redistributing vorticity into the gyre's interior.

In conclusion, while wind stress is important in the subtropics, it is less important in the overall subpolar gyre vorticity balance (see the total vorticity balance in Fig.~\ref{fig:ctrl-bvb-subpolar}b, consistent with \citeA{Yeager2015} and \citeA{Corre2020}).
The subpolar gyre strongly deviates from the topographic-Sverdrup balance since the non-linear advection and viscosity terms also play a key role in setting the flow.
Flow-topography interactions are important in steering the flow; bottom pressure anomalies are responsible for steering the interior circulation, while bottom drag guides the western boundary flow, especially near continental shelves, due to opposite-signed vorticity input as the wind stress curl.
Non-linear advection and viscosity also influence the circulation by redistributing vorticity around the gyre and widening thin boundary currents.
The secondary role of the wind stress suggests that other surface forcing mechanisms (e.g., surface buoyancy forcing \cite{Delworth2016}) that can affect non-linear advection, lateral and bottom friction, and bottom pressure anomalies may be important in setting the subpolar gyre circulation.

\section{Results: Response of the gyres to the NAO}
\label{section:perturbations-bvb-analysis}
In this section, we investigate the physical processes that govern how the North Atlantic gyre responds to NAO variability.
We run four experiments each for 100 years where we alter the surface forcing variables to mimic an NAO index of $-2$, $-1$, $+1$, and $+2$ times the standard deviation observed in the yearly-mean NAO state in Fig.~\ref{fig:expt_setup}b.

\subsection{Subtropical gyre}

\begin{figure}[ht]
    \includegraphics[width=1.0\textwidth]{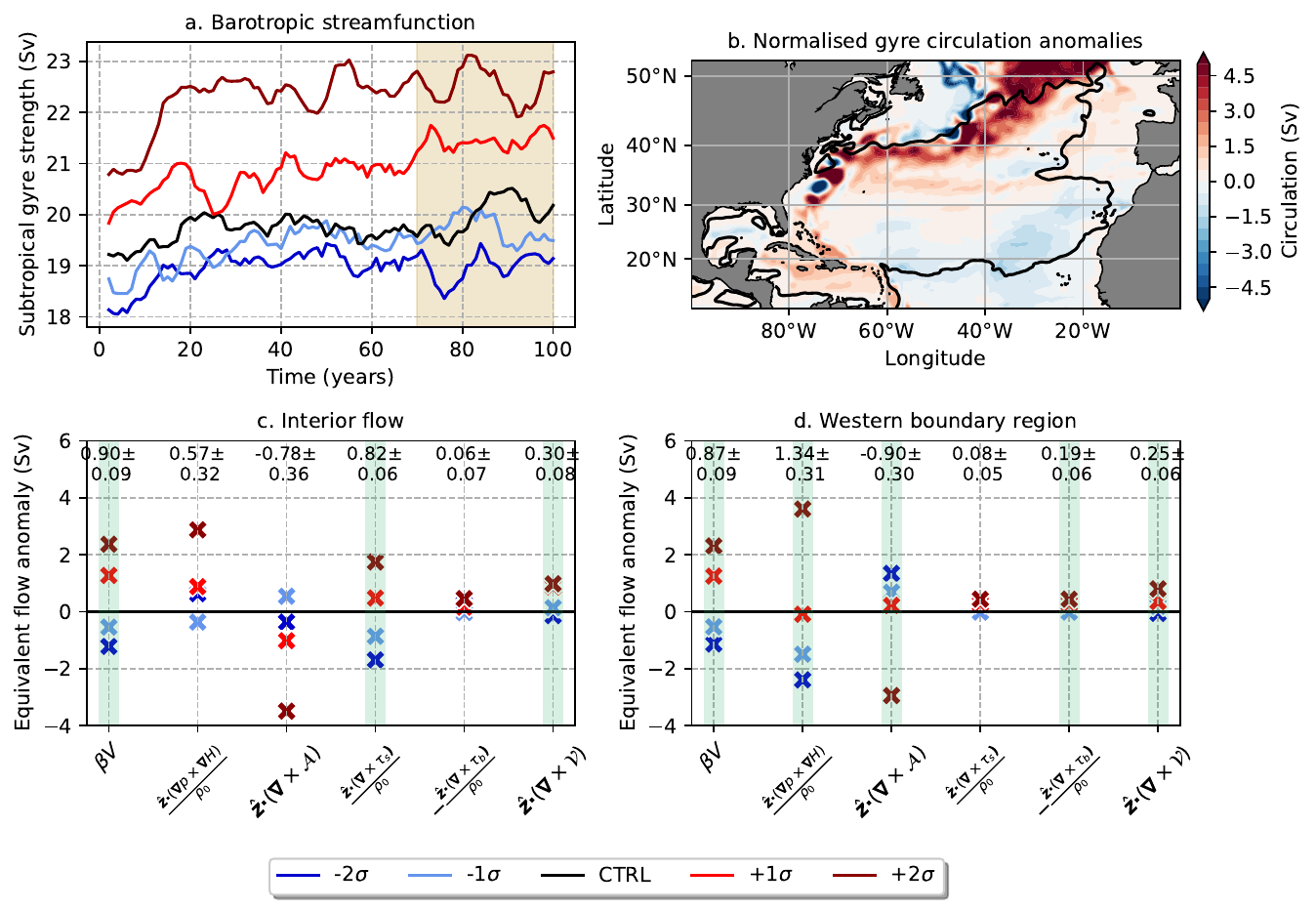}\vspace{-1em}
    \centering
    \caption{(a).~Time series of the subtropical gyre strength estimated using the 95$^{\textrm{th}}$ percentile method with a 5-year rolling mean. The orange shading represents the time period used to analyze the time-mean vorticity budget.
    (b).~Linearly regressed anomalies in the gyre circulation with the NAO index. Black contour indicates $\Psi_{\rm{base}} = 0~\rm{Sv}$.
    (c).~Equivalent flow anomaly in the gyre's interior due to vorticity induced by leading order terms in the budget. Each variable is linearly regressed against the NAO anomaly, as shown at the top of each column along with the corresponding standard errors (in Sv per unit NAO index). Green bars denote that the linear regression is statistically significant ($p < 0.1$). Colors of the crosses correspond to the legend in panel~(a).
    (d).~Same as panel~c, but for the western boundary region.}
    \label{fig:subtropical-bvb-perturbations}
\end{figure}

The subtropical gyre quickly adjusts to the NAO surface forcing perturbations (Fig.~\ref{fig:subtropical-bvb-perturbations}a).
The circulation intensifies under positive NAO forcing anomalies and weakens under negative NAO forcing anomalies relative to the control.
This adjustment is spatially variable (Fig.~\ref{fig:subtropical-bvb-perturbations}b) due to the differing dynamical balances operating in different regions (outlined in section~\ref{section:control-bvb-analysis}).
For example, we observe larger circulation anomalies in the western boundary compared to the interior as the NAO shifts from negative to positive phases (Fig. \ref{fig:subtropical-bvb-perturbations}b).
We perform a linear regression analysis for all leading order terms in Eq.~\eqref{eq:gyre-strength-bvb} with the NAO index (shown by the numbers at the top of Figs.~\ref{fig:subtropical-bvb-perturbations}c-d) to understand whether individual terms in the budget also respond linearly with the NAO index.

Since the interior meridional flow has a negative sign, we multiply the rewritten vorticity budget (Eq.~\eqref{eq:gyre-strength-bvb}) by $-1$ to ensure that positive anomalies signify an intensified gyre circulation.
In the previous section, we showed that the interior of the North Atlantic subtropical gyre follows a topographic-Sverdrup balance.
In the perturbation experiments, we observe stronger anticyclonic vorticity input by the wind stress for positive NAO phase experiments and weaker input for negative NAO phase experiments (Fig.~\ref{fig:subtropical-bvb-perturbations}c).
The regression numbers at the top of Fig.~\ref{fig:subtropical-bvb-perturbations}c convey that for a unit increment in the NAO index, the wind stress curl provides an additional anticyclonic vorticity equivalent to $0.82$~Sv, which drives about 91\% of the anomalous gyre transport (note that this does not imply other processes are negligible as there is compensation, see Fig.~\ref{fig:subtropical-bvb-perturbations}c).
The dominance of wind stress can alternatively be quantified from the quick adjustment timescales of the gyre to NAO forcing (see \cite{Anderson1975} for a investigation of the ocean circulation's adjustment processes and time scales due to wind stress; Fig.~\ref{fig:subtropical-bvb-perturbations}a).
A linear regression analysis similar to that carried here but for the first decade (figure not shown) shows that $0.84\,\textrm{Sv}$ circulation anomalies occur within the first 10 years.
Thus, the NAO-induced wind stress variability plays a central role in generating circulation anomalies in the subtropical gyre's interior.

The bottom pressure torque and non-linear advection curl change non-linearly with the NAO index across experiments (Fig.~\ref{fig:subtropical-bvb-perturbations}c), however, these anomalies cancel each other to a large extent.
The bottom drag curl shows a slight positive trend with the NAO index.
However, this trend cannot be considered robustly linear, since the standard error is of the same magnitude as the trend.
Finally, the curl of viscosity also increases as the NAO index becomes more positive, which could be due to a number of processes such as changes in near-surface turbulence led by wind stress and changes in stratification.

The western boundary region in our control simulation exhibited a complex balance, with major roles for the bottom drag curl, non-linear advection curl, and the curl of viscosity.
Therefore, it is not surprising that all these processes also play a role in the gyres response to NAO forcing (Fig.~\ref{fig:subtropical-bvb-perturbations}d).
However, it is important to emphasize that these processes are simply a balance to ensure anomalous flow, which is ultimately driven by NAO-induced variations in the surface forcing.
More specifically, it is the wind stress curl that inputs anticyclonic vorticity on the leading order that in turn changes the interior flow, and these changes are subsequently reflected in the return flow, that is, the western boundary current. Other processes, such as NAO-induced changes in the southward advection of the North Atlantic Deep Water have also been shown to influence the gyre circulation \cite{Kim2024}, but our analysis indicates that the effect of these changes on the circulation are of second order when compared with the wind stress curl.

In the western boundary region, the bottom pressure torque explains vorticity anomalies equivalent to $1.34~$Sv per unit NAO index in the western boundary.
Interestingly, the gyre circulation intensifies only by $0.87$~Sv per unit NAO index, suggesting that the bottom pressure torque is more sensitive to NAO variability than gyre circulation.
About two-thirds of the bottom pressure torque anomalies are compensated by the non-linear advection term (Fig.~\ref{fig:subtropical-bvb-perturbations}d), implying that the flow has a tendency to redistribute anomalous vorticity generated by the bottom pressure variations.
The bottom drag curl also increases as we transition from negative to positive NAO phase experiments, which may be due to an increase in the gyre's abyssal flow linked to its stronger barotropic nature or due to changes in the flow speed.
Additionally, stronger western boundary currents in a positive NAO phase is often associated with intensified shear due to large horizontal velocity gradients.
The increased gradients likely enhance the viscous dissipation of vorticity in these experiments ($0.25$~Sv increase per unit NAO index).

\subsection{Subpolar gyre}

\begin{figure}[ht]
    \includegraphics[width=1.0\textwidth]{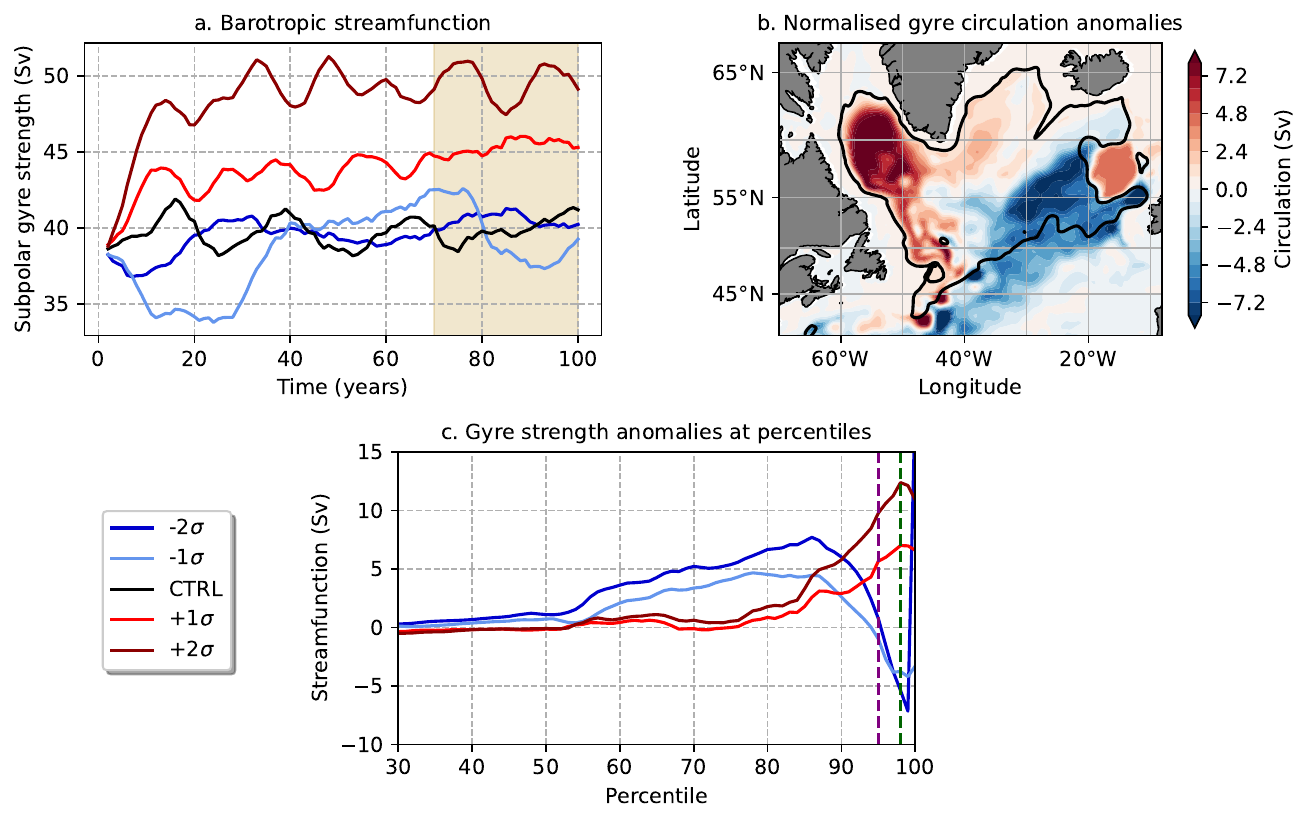}\vspace{-1em}
    \centering
    \caption{(a).~Time series of the subpolar gyre strength estimated using the 95$^{\textrm{th}}$ percentile method with a 5-year rolling mean. The orange shading represents the time period used to analyze the time-mean vorticity budget.
    (b).~Linearly regressed anomalies in the subpolar gyre circulation with the NAO index. Black contour indicates $\Psi_{\rm{base}} = -10~\rm{Sv}$ obtained by averaging the barotropic streamfunction for the CTRL experiment over the last 30 years.
    (c).~Distribution of subpolar gyre's streamfunction anomalies (with respect to CTRL) with percentile. Vertical dashed lines are plotted for $\Psi_{\rm{gyre}}$ (purple) and the 98$^{\rm{th}}$ percentiles (green).}
    \label{fig:subpolar-perturbations-intro}
\end{figure}

The subpolar gyre responds non-linearly to changes in the NAO index: strengthening for positive NAO phase experiments while fluctuating around the control for negative NAO phase experiments (Fig.~\ref{fig:subpolar-perturbations-intro}a).
This non-linear behavior is linked to the opposing trends in the circulation with the NAO index between the western boundary and interior regions (Fig.~\ref{fig:subpolar-perturbations-intro}b), which differs from a single-signed circulatory response in the analysis by \citeA{Kim2020}, who considered only the impact of NAO-induced surface heat flux anomalies, not winds, on the gyre circulation.
Specifically, the circulation in the Labrador and part of the Irminger Seas intensifies as we transition from negative to positive NAO phase experiments.
On the other hand, the eastern subpolar gyre's (interior flow's) circulation reduces with increments in the NAO index.
This inverse trend of the interior flow with the NAO index places a larger number of grid cells between the $50^{\rm{th}}$ and $95^{\rm{th}}$ percentiles in the negative NAO phase experiments (Fig.~\ref{fig:subpolar-perturbations-intro}c).
Choosing a higher percentile value (e.g., $98^{\rm{th}}$ denoted by the green dashed line in Fig.~\ref{fig:subpolar-perturbations-intro}c) leads to a more symmetric gyre circulation response to the NAO index while maintaining a vorticity balance similar to that of the $95^{\rm{th}}$ percentile (Fig.~S2b-c).
However, in this case the resulting western boundary regions fail to adequately capture the Labrador Current (Fig.~S2d-g).
Moreover, larger percentiles could be associated with circulation due to standing eddies rather than the gyre's overall cyclonic circulation.
Therefore, we retain the $95^{\rm{th}}$ percentile to define our gyre strength.
We first analyze the gyre's interior flow and western boundary region using Eq.~\eqref{eq:gyre-strength-bvb}, followed by an explanation for the dipole circulation anomalies in Fig.~\ref{fig:subpolar-perturbations-intro}b.

\begin{figure}[h!]
    \includegraphics[trim={20cm 0cm 19.5cm 0cm}, clip, width=0.78\textwidth]{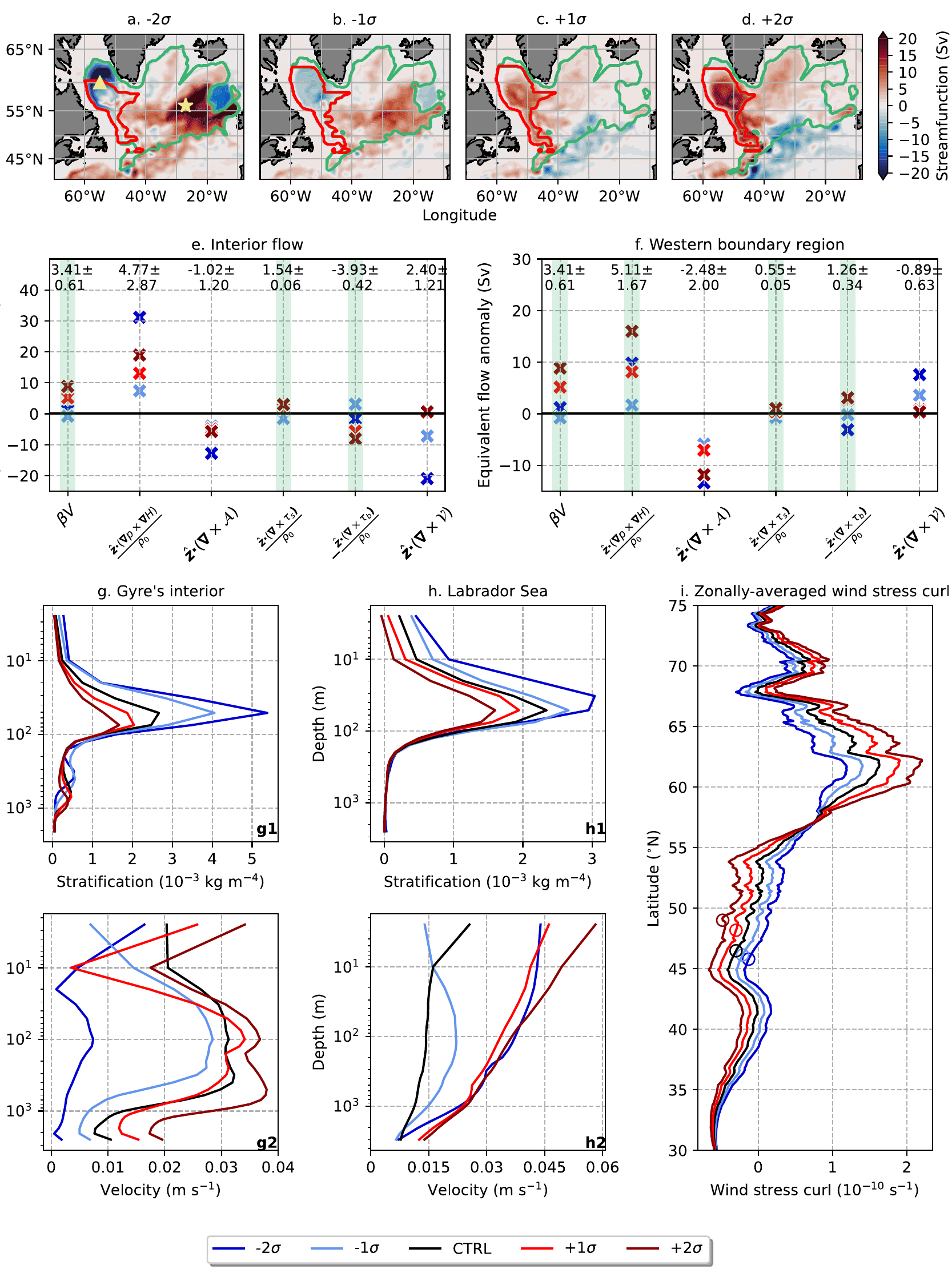}\vspace{-1em}
    \centering
    \caption{
    (a-d).~Gyre circulation anomalies for perturbation experiments. Red and green contours represent the western boundary and the interior regions respectively. The light yellow markers indicate locations for which stratification and horizontal velocity profiles are plotted in panels~(g) and~(h).
    (e).~Equivalent flow anomaly in the gyre's interior due to vorticity induced by leading order terms in the budget. Each variable is linearly regressed against the NAO anomaly, as shown at the top of each column along with the corresponding standard errors (in Sv per unit NAO index). The $-2\sigma$ experiment is not included in the regression analysis. Green bars represent variables for which the linear regression is statistically significant ($p < 0.1$). Colors of the crosses correspond to the center-bottom legend.
    (f).~Same as panel~(e), but for the western boundary.
    (g).~Stratification (g1) and horizontal velocity (=$|\bu|$; g2) profiles for a region in the gyre's interior (marked by the light yellow star in panel~(a)). We define stratification as the vertical gradient of potential density referenced to $2000\,\textrm{dbar}$.
    (h).~Same as panel (g), but for a location in the western boundary region marked by the light yellow triangle in panel~(a).
    (i).~Meridional variation in the zonally-averaged wind stress curl in the North Atlantic. Circles represent the latitude of zero streamfunction at 30$^{\circ}$W, which we define as the North Atlantic Current latitude.
    }
    \label{fig:subpolar-bvb-perturbations}
\end{figure}

We notice differences in the area of integration between $-2\sigma$ and other perturbation experiments (Figs.~\ref{fig:subpolar-bvb-perturbations}a-d); such differences were negligible in the subtropical gyre.
As a result, we omit the $-2\sigma$ experiment from the regression analysis used to derive the numbers at the top of Figs.~\ref{fig:subpolar-bvb-perturbations}e-f.
Bearing in mind the exclusion of the $-2\sigma$ experiment, the interior flow changes by $3.41~\rm{Sv}$ per unit NAO index due to the combined effect of all physical processes in the vorticity budget.
If the gyre were completely Sverdrupian, the gyre strength would have increased by $1.54\,\textrm{Sv}$ for each increment in the NAO index (fourth column in Fig.~\ref{fig:subpolar-bvb-perturbations}e).
Assuming that most of the gyre's circulation adjusts to the NAO-driven wind stress anomalies within a few years, a change of $2.50\,\mathrm{Sv}$ per unit NAO index within the first 10 years further supports the smaller contribution of wind stress compared to the subtropical gyre.
Thus, unlike in the subtropical gyre, NAO-induced wind stress anomalies do not explain a large fraction of the subpolar gyre variability, highlighting the role of other processes in steering the gyre circulation, such as surface buoyancy forcing (see e.g., \citeA{Kim2020, Kim2024}).

The bottom pressure torque inputs a large amount of cyclonic vorticity in the interior (significantly larger than the overall change in the circulation strength).
Increased surface heat loss in the positive NAO phase experiments in the gyre's interior reduces stratification (Fig.~\ref{fig:subpolar-bvb-perturbations}g1) that promotes stronger circulation throughout the water column (Fig.~\ref{fig:subpolar-bvb-perturbations}g2).
The enhanced deep ocean circulation is more strongly steered by bottom topography, and the excess vorticity represented by the bottom pressure torque may be redistributed towards other regions by the curl of non-linear advection \cite{Stewart2021}.
On the other hand, we observe reduced bottom pressure torque in most of the subpolar gyre's interior in the negative NAO phase experiments attributed to to reduced surface heat loss, except near the Rockall Trough (not shown).
The intensified interior flow in the negative NAO phase experiments relative to the control (Fig.~\ref{fig:subpolar-bvb-perturbations}a-b) significantly intensifies flow-topography interactions near the Rockall Trough, which increases the net bottom pressure torque in the interior.
Hence, we observe positive anomalies in the area-integrated bottom pressure torque for all perturbation experiments (Fig.~\ref{fig:subpolar-bvb-perturbations}e).

The large input of cyclonic vorticity by the bottom pressure torque is compensated by other processes. The bottom drag curl counteracts NAO-induced change in the circulation (fifth column in Fig.~\ref{fig:subpolar-bvb-perturbations}e), thereby reflecting its tendency to oppose the vorticity input induced by wind stress curl \cite{Stewart2021}.
Enhanced flow-topography interactions due to reduced stratification may promote vorticity input due to bottom drag in the positive NAO phase experiments; the opposite being true for the $-1\sigma$ NAO phase experiment (although the $-2\sigma$ experiment where the interior flow's boundary deviates significantly from other experiments does not follow this trend).
The curl of viscosity inputs small vorticity anomalies except in the $-2\sigma$ experiment, where thin boundary currents in the Labrador Sea (area of integration is noticeably different between $-2\sigma$ and other experiments) and opposing circulatory flows near the Rockall Trough generate stronger horizontal shear.

We now shift our focus to the western boundary region, where the circulation changes relatively linearly with the NAO index (compare regions bounded by the red contours in Fig.~\ref{fig:subpolar-bvb-perturbations}a-d).
Similar to the interior, there is significant compensation between different physical processes.
Here, the bottom pressure torque explains $5.11$~Sv equivalent flow for a unit increment in the NAO index (Fig.~\ref{fig:subpolar-bvb-perturbations}f), compared to the $3.41$~Sv change observed in the subpolar gyre circulation.
The increase in bottom pressure torque as the NAO phase becomes more positive can be attributed to enhanced surface heat loss in the Labrador and Irminger Seas, which reduces stratification and promotes flow-topography interactions (Fig.~\ref{fig:subpolar-bvb-perturbations}h1).
However, it is interesting that even though the stratification changes linearly with the NAO index, the flow anomalies are non-linearly related to the index (Fig.~\ref{fig:subpolar-bvb-perturbations}h2).
The isopycnal restructuring observed here is similar to that shown in the surface heat flux only perturbations due to the NAO by \citeA{Kim2024}, who infer that both NAO-induced anomalous as well as background surface heat fluxes exert a prominent control on the subpolar gyre circulation.
We suspect that these changes in the surface buoyancy forcing and stratification play a major role in changing the AMOC strength by $1.17 \pm 0.13\,\mathrm{Sv}$ per unit NAO index (figure not shown).
Moreover, the AMOC shows a similar trend (in time as well as between experiments) as the subpolar gyre circulation, consistent with past studies that suggested that the two circulatory features are coupled \cite{Lohmann2009, Yeager2020}.
However, a full analysis of the NAO impacts on the AMOC is outside the scope of our study.

Non-linear advection redistributes about 50\% of the anomalous vorticity input by the bottom pressure torque either towards the continental shelf or the interior.
The bottom drag curl provides about one-third of the gyre circulation anomalies due to NAO forcing (fifth column in Fig.~\ref{fig:subpolar-bvb-perturbations}f).
The $-2\sigma$ experiment shows the opposite response: the bottom drag curl damps the circulation anomalies, which may be due to differences in the western boundary region with other experiments.
The increase in bottom drag curl as the NAO index becomes more positive can be attributed to changes in both winds and surface buoyancy forcing.
First, bottom drag curl has a tendency to counteract the large-scale cyclonic vorticity provided by the wind stress curl in the subpolar gyre.
Since the wind stress curl increases as we transition from negative to positive NAO phase experiments, so does the bottom drag curl.
Second, stratification in the western boundary region increases in negative NAO phase experiments and reduces in positive NAO phase experiments due to variations in surface heat loss (Fig.~\ref{fig:subpolar-bvb-perturbations}g).
Reduced stratification promotes flow-topography interactions, which may cause a more pronounced role of the bottom drag curl in positive NAO phase experiments.
Lastly, the local wind stress curl unsurprisingly plays an insignificant role in driving the western boundary flow.
%

We now address the cause behind the inverse relationship of the strength of subpolar gyre's interior with the NAO index (Figs.~\ref{fig:subpolar-bvb-perturbations}a-d).
We hypothesize that this inverse relationship is linked to non-local anomalies in the wind stress curl further south that are advected towards the subpolar gyre's interior.
More specifically, meridional shifts in the zero wind stress curl latitude (Fig.~\ref{fig:subpolar-bvb-perturbations}i) contribute to a northward displacement of the Gulf Stream for positive NAO phase experiments and southward for negative NAO phase experiments, consistent with past studies \cite{Marshall2001-NAO, Wolfe2019, Silver2021}.
The Gulf Stream feeds the North Atlantic Current, whose pathway changes in a similar manner -- the flow is more zonal and limited to lower latitudes for negative NAO phase experiments (compare circles in Fig.~\ref{fig:subpolar-bvb-perturbations}i), while the North Atlantic Current is more meridional for positive NAO phase experiments.
The wind stress curl in the negative NAO phase experiments is relatively small at the latitude of the North Atlantic Current (blue lines in Fig.~\ref{fig:subpolar-bvb-perturbations}i), and thus the North Atlantic Current flows zonally towards the eastern subpolar gyre to strengthen the interior circulation.
In contrast, in the positive NAO phase experiments the North Atlantic Current is more strongly directed equatorward by the anticyclonic wind stress curl (red lines in Fig.~\ref{fig:subpolar-bvb-perturbations}i) towards the subtropical gyre's interior.

In conclusion, the interior flow in the subpolar gyre is influenced by both local and remote processes, with large compensations between opposing terms.
NAO-induced anomalies in the subpolar gyre's interior are governed by a combination of wind stress, stratification (influenced by surface buoyancy forcing as well as non-linear advection), and flow-topographic interactions.
In the western boundary, local wind stress curl plays a negligible role in driving NAO-induced circulation anomalies, which is supported by topography primarily by the bottom pressure torque (150\%) and secondarily by the bottom drag curl (37\%).
Surface buoyancy forcing plays a strong role in influencing these flow-topographic interactions in the western boundary through changes in stratification, consistent with \citeA{Delworth2016} and \citeA{Khatri2022}.

\section{Summary and Discussion}
\label{section:summary}

In this paper, we examined the role of a range of ocean dynamical processes impacting the equilibrium response of the North Atlantic gyre circulation to surface forcing anomalies from the North Atlantic Oscillation (NAO; see Fig.~\ref{fig:expt_setup}a).
By appropriate choices of area-integration, we derived a barotropic vorticity budget (Eq.~\eqref{eq:gyre-strength-bvb}) that explicitly relates vorticity sources/sinks, such as wind stress curl, bottom pressure torque and non-linear advection, directly and quantitatively to a scalar metric of gyre strength.
We analyzed this budget within a series of global ocean-sea ice model experiments at~$0.25^{\circ}$ horizontal resolution (described in Fig.~\ref{fig:expt_setup}a) forced by idealized time-invariant perturbation patterns representing positive and negative NAO phases to understand the mechanisms through which the NAO impacts the subpolar and subtropical gyres.

In the control experiment, the subtropical gyre's interior follows topographic-Sverdrup balance (Eq.~\eqref{eq:topo-Sverdrup-balance}; Fig.~\ref{fig:ctrl-bvb-subtropics}b), since other processes such as the curl of non-linear advection, viscosity, and bottom drag terms play a minor role.
The interior flow strengthens by $0.90$~Sv for every unit increment in the NAO index, of which $0.82$~Sv (or~$91$\%) is explained by wind stress curl anomalies (Fig.~\ref{fig:subtropical-bvb-perturbations}c).
On the other hand, the western boundary current flow is primarily guided locally by the bottom pressure torque and secondarily by the bottom drag curl \cite{Hughes2001}.
Overall, the NAO-induced circulation anomalies in the western boundary (Fig.~\ref{fig:subtropical-bvb-perturbations}d) are balanced by a combination of bottom pressure torque ($154$\%), non-linear advection curl ($-103$\%), viscosity curl ($29$\%), and bottom drag curl ($22$\%).
More than half of the NAO-induced vorticity anomalies due to the bottom pressure torque anomalies are redistributed towards the interior or the continental shelves by the non-linear advection curl (Fig.~\ref{fig:subtropical-bvb-perturbations}d).
Thus, the NAO drives variations in the subtropical gyre through wind stress which are then modified by flow-topography interactions.
Our results suggest a minimal role for surface buoyancy forcing in modulating the subtropical gyre due to NAO forcing.

In contrast to the subtropical gyre, the subpolar gyre circulation responds non-linearly to the NAO forcing.
While the circulation intensifies for positive NAO phase experiments, it fluctuates around the control for negative NAO phase experiments.
We note that this non-linear trend of the subpolar gyre circulation strength with the NAO index is dependent on our metric of gyre strength (Fig.~\ref{fig:subpolar-perturbations-intro}c).
Unlike in the subtropical gyre, the wind stress curl does not dominate the NAO-induced anomalies in the subpolar gyre,
suggesting an enhanced role for other mechanisms.
Although the bottom pressure torque contributes noticeably to the circulation, the subpolar gyre deviates from the topographic-Sverdrup balance since viscosity, bottom drag curl and non-linear advection noticeably affect the circulation (Fig.~\ref{fig:ctrl-bvb-subpolar}b).
In addition, non-local anomalies in the wind stress curl (i.e., arising from latitudes further south) also impact the long-term response of the gyre's interior, and lead to an inverse trend in the circulation with the NAO index (Fig.~\ref{fig:subpolar-perturbations-intro}b).
Similar to the subtropical gyre, the western boundary flow is locally steered primarily by the bottom pressure torque and secondarily by the bottom drag curl \cite{Hughes2001}.
The role of topography is further exemplified in the positive NAO phase experiments, where surface cooling reduces stratification and thus promotes flow-topography interactions.
Moreover, about half of the NAO-induced bottom pressure torque anomalies are balanced by the non-linear advection curl (Fig.~\ref{fig:subpolar-bvb-perturbations}f).
We demonstrate that the NAO-induced anomalies in the subpolar gyre circulation is strongly impacted by variations in the stratification (controlled by surface buoyancy forcing), flow-topography interactions, and the North Atlantic Current.
These results highlight the importance of both winds and surface buoyancy forcing in steering NAO-induced variations in the subpolar gyre circulation.

Although the impact of winds in driving the gyre circulation is well-known \cite{Sverdrup1947}, the role of surface buoyancy forcing has only recently been highlighted \cite{Hogg2020, Bhagtani2023}.
The barotropic vorticity budget used in this analysis does not contain an explicit term that quantifies surface buoyancy forcing of the gyre circulation.
In other words, we cannot directly compare the importance of winds and surface buoyancy forcing in steering the large-scale gyre circulation only using the barotropic vorticity budget.
Modifications to our model setup (e.g., running simulations with one surface forcing change at one time) in the future can help to cleanly separate the contributions of wind stress and surface buoyancy forcing.
Nevertheless, the impact of surface buoyancy forcing can be understood through further analysis; for example, changes in stratification (see Figs.~\ref{fig:subpolar-bvb-perturbations}g-h) and adjustment timescales, resulting in modulation via other processes in the vorticity budget such as bottom pressure torque and bottom drag curl.

We use constant amplitude NAO-induced surface forcing perturbations throughout the year, since our focus in this paper was to understand the long-term response of the North Atlantic gyre circulation.
Using a seasonally-varying NAO forcing (or alternatively, a winter-only NAO forcing) may change the relative contribution of momentum- and buoyancy-induced changes in the circulation (since the multi-year baroclinic adjustment timescales could be altered by the quick barotropic response).
Future work could investigate whether seasonally-varying NAO forcing leads to different gyre responses, particularly by isolating winter-specific forcing anomalies.
A comparison with studies that use persistent winter NAO indices \cite{Eden2001, Lohmann2009, Kim2024} suggests that the key dynamical processes identified here (e.g., changes in stratification, meridional displacement of the North Atlantic Current, or the influence of the wind stress curl) remain robust, which instills confidence in the robustness of the gyres' force balance.
Our simulations are run at a $0.25^{\circ}$ horizontal resolution, which is unable to fully resolve baroclinic modes at high latitudes.
While our model's large-scale gyre balances are similar to other higher resolution analyses \cite{Yeager2015, Corre2020}, we observe a larger contribution of the viscous curl term in steering the subpolar gyre, which may be related to a strong spurious role for viscosity in diffusing thin boundary currents in the subpolar gyre.
Regional configuration simulations (e.g., see \citeA{Barnes2024} and references therein) could be used to enhance horizontal resolution using similar computational resources.
Alternatively, the use of more sophisticated subgrid scale parameterizations that transfer energy through backscatter or stochastic modeling (e.g., see \citeA{Yankovsky2024}) may also more realistically represent the curl of viscosity term.
Finally, our analysis is performed using an ocean--sea ice model, therefore the feedback of NAO-induced changes in the oceanic state (e.g., sea surface temperature) on the atmosphere is not considered.

The NAO index is expected to become more positive in the future due to increased greenhouse concentrations and Arctic sea ice loss \cite{Folland2009, Gillett2013}.
Our results suggest that both gyres will strengthen for long-term positive NAO phases, thus, we expect gyres to contribute more towards poleward heat transfer under positive NAO phases \cite{Bhagtani2024}.
An interesting result is the asymmetric change in the gyre circulation with the NAO state (Figs.~\ref{fig:subtropical-bvb-perturbations}a and~\ref{fig:subpolar-perturbations-intro}a), which may exist due to several processes, including long-term memory in the ocean.
For instance, \citeA{Lohmann2009} also found an asymmetric subpolar gyre response to the NAO: initial strengthening for positive phases followed by weakening at equilibrium, while a gradual weakening for negative phases.
The NAO also affects the meridional overturning circulation on long timescales due to variations in dense water formation in the Labrador and Irminger Seas.
In our experiments, the meridional overturning circulation intensified with increments in the NAO index (not shown), consistent with past studies \cite{Delworth2016}.
Rewriting a depth-dependent vorticity budget to explicitly include the overturning strength (e.g., by decomposing into a poleward and equatorward flow at the depth of maximum vertical velocity) could be explored to better understand physical processes driving the meridional overturning circulation.
Furthermore, analyzing the individual responses of these circulatory features using different partition techniques \cite{Ferrari2011, Jones2023} could be useful in predicting how the ocean's heat transport will change on sub-seasonal to multi-decadal timescales due to the NAO.

\section{Open Research}
Python code used for generating figures is available at \href{https://github.com/dhruvbhagtani/NAO-vorticity-budget-gyres}{github.com/dhruvbhagtani/NAO-vorticity-budget-gyres}.
Preprocessed outputs along with a copy of the jupyter notebooks used to reproduce figures is available at \href{https://doi.org/10.5281/zenodo.15092440}{10.5281/zenodo.15092440}.
Source code for MOM6 can be accessed at \href{https://github.com/NOAA-GFDL/MOM6}{github.com/NOAA-GFDL/MOM6}.

\acknowledgments
D.B.~would like to thank Kai Kornhuber and Isla Simpson for discussions on the North Atlantic Oscillation and its impact on the large-scale ocean circulation.
D.B.~also expresses gratitude to Brandon Reichl for their help with running MOM6 and Julia Neme for useful inputs on dynamically motivated area-integrations for gyres.
We thank Christopher Bladwell for useful suggestions on a draft of this manuscript.
We would also like to thank the vibrant community of the Consortium for Ocean--Sea Ice Modeling in Australia (\href{https://cosima.org.au}{cosima.org.au}) for fruitful discussions.
We also thank the editor along with the two reviewers that helped to improve the quality of this manuscript.
Our analyses were facilitated with the Python packages \texttt{dask} \cite{Rocklin2015}, \texttt{eofs} \cite{Dawson2016}, \texttt{gcm-filters} \cite{Loose2022}, and \texttt{xarray} \cite{Hoyer2017}.
Computational resources were provided by the National Computational Infrastructure at the Australian National University, which is supported by the Commonwealth Government of Australia.
D.B.~and~A.M.H.~were supported by the Australian Research Council Center of Excellence for Climate Extremes CE170100023.
N.C.C.~acknowledges funding from the Australian Research Council under DECRA Fellowship DE210100749.


%

%




%
%
%
%
%

\end{document}